\documentclass[doublecol,amsmath,amssymb,cite]{epl2}

\title{Jamming and Geometry of Two-Dimensional Foams}

\shorttitle{} 

\author{Gijs Katgert\inst{1,2}  \and Martin van Hecke\inst{1}}
\shortauthor{G. Katgert \etal}

\institute{
\inst{1}
  Kamerlingh Onnes Laboratorium, Universiteit Leiden,
   P.O. Box 9504, 2300 RA Leiden, The Netherlands. \\
\inst{2}   
  School of Physics, The University of Edinburgh, James Clerk Maxwell Building, The Kings Buildings,   		 
   Mayfield Road, Edinburgh EH9 3JZ, United Kingdom. \\}


\abstract{We experimentally probe the vicinity of the jamming point J, located at a density $\phi$ corresponding to random close packing ($\phi_{rcp} = 0.842$), in two dimensional, bidisperse packings of foam bubbles. We vary the density of the foam layer and extract geometrical measures by image analysis.
We confirm the predicted scaling of the average contact number Z with $\phi$ and compare the distribution of local contact numbers to a simple model. We further establish that the distribution of areas $p(A)$ strongly depends on $\phi$. Finally, we find that the distribution of contact forces $p(f)$ systematically varies with density.}

\begin{document}

\maketitle
Since the seminal work by Bolton and Weaire \cite{bolton} soft frictionless discs or spheres have become the Ising model for the Jamming transition \cite{durian, ohern, vanhecke}. Jamming is believed to capture the transition between flow and arrest in a wide variety of disordered media such as foams, emulsions, granular media and (colloidal) suspensions, and this idea has lead to an upsurge of simulations  revealing critical behaviour as a function of the distance to the critical point J, which is reached precisely when the applied pressure vanishes \cite{ohern, olsson, tighe}. While some of these predictions have been tested experimentally \cite{mason, st-jalmes,majmudar,nordstrom}, many others still await experimental verification.
	
While the original incarnations of the soft sphere model explicitly make the link to foams \cite{bolton, durian}, this connection has not been explored in recent times. Nevertheless, foam bubbles, as well as emulsion droplets are the closest physical analogue of frictionless spheres. Their elastic interaction is close to that of a linear spring \cite{lacasse, morse} and solid friction is absent. Under flow, only a velocity dependent viscous friction, that is now fairly well understood \cite{olsson,tighe,katgert,denkov} acts on the bubbles.

In this Letter we experimentally probe the behaviour of jammed packings near point J. We do this by generating two-dimensional packings of foam bubbles. We vary the packing fraction, $\phi$, and for each density generate many distinct static packings of foam bubbles. We then extract various statistical and topological quantities through image analysis and investigate whether these measures signal an approach of the jamming transition at $\phi_c$ as the density is varied.

First, we investigate the scaling of the average contact number per bubble Z with the distance to $\phi_c$. Our central result is that the contact number scales like ${\rm Z-Z}_c \sim (\phi-\phi_c) ^{0.5}$. To this end, we first resolve the apparent discrepancy between our findings and simulations by pointing out the differing ways of measuring $\phi$ between simulation and experiment. We find $\phi_c$ to be located around $\phi_c =0.84$, in excellent agreement with previous predictions \cite{bolton,ohern,olsson,lechenault}, and obtain, for the first time, a quantitative experimental confirmation of critical scaling at point J in an appropriate experimental system.

Besides this global measure, we also investigate the distribution of contact numbers at the bubble scale as a function of global Z. We compare to a recent model \cite{henkes} and find excellent agreement with no adjustable parameters, which is remarkable given that this theory was developed for frictional packings.

The distribution of free available area per bubble, $p(A)$ plays an important role in statistical mechanics descriptions of jammed matter \cite{makse, aste, lechenaultII}. We show that the correct way of extracting $p(A)$ is by tesselating the foam packing with the navigation map \cite{medvedev,corwin}. We fit the obtained distributions of free area per bubble to gamma distributions \cite{aste,kermode}, and show that, in contrast to granular packs, this distribution is far from universal: as we decrease the packing fraction $\phi$ towards point J, we find an excess of large available areas, and in conjunction with this an increase of the compactivity.

Finally, our navigation map tiling allows us to extract the deformed interfaces between adjacent bubbles. These facets are proportional to the contact forces, and we thus extract the distribution of the contact forces $p(f)$. We find that its tail is close to exponential for $\phi$ close to $\phi_c$, and becomes steeper for large $\phi$. \cite{brian}.

\begin{figure}[t]
\begin{center}
\includegraphics[width=\columnwidth]{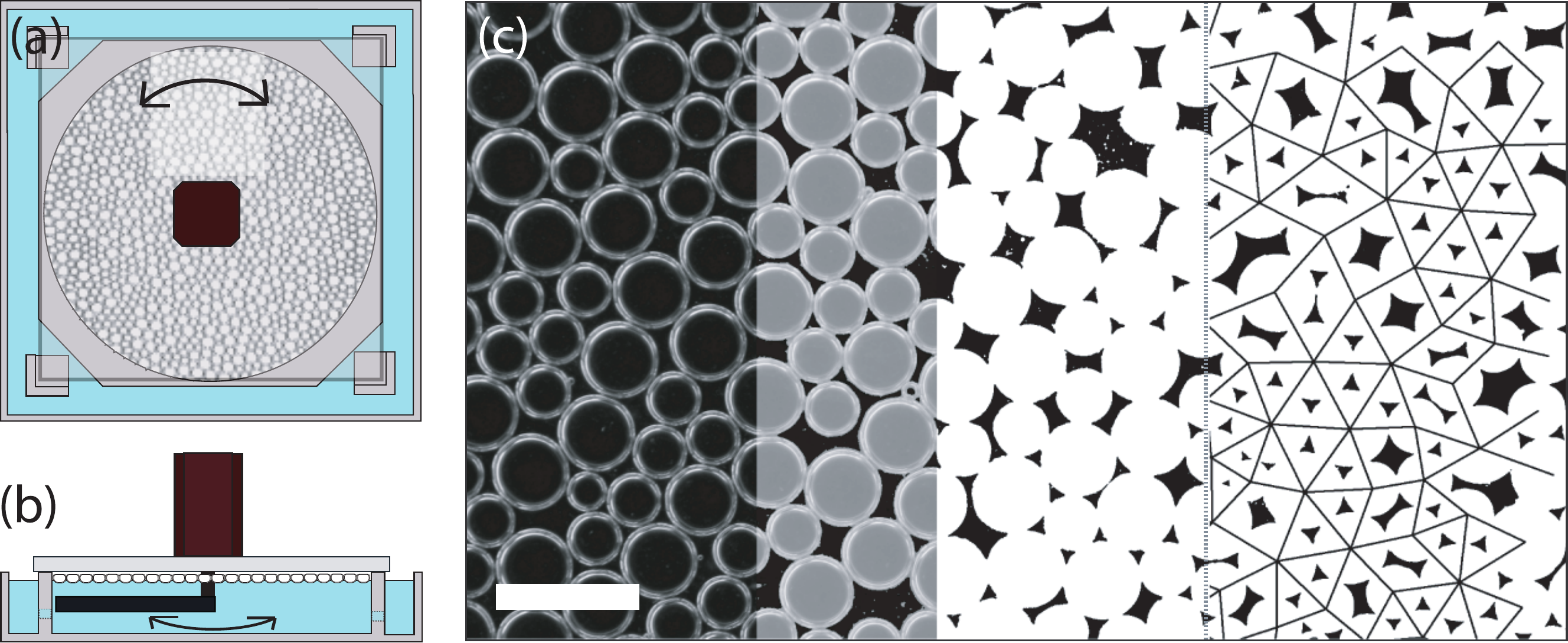}
\caption{(Color online)
Top (a) and side (b) view of the setup. (a) Bubbles are contained in a circular reservoir and covered by a glass plate. Images are taken in the highlighted area. (b) A stepper motor drives a rod that stirs the bulk solution, thereby rearranging the packing. (c) From left to right: raw image, raw image with reconstructed bubble areas, reconstructed bubble areas, from which $\phi$ is calculated, and contact network, from which Z is found. Scalebar denotes 5 mm.} \label{setup}
\end{center}
\end{figure}
\emph{Setup ---}  We prepare a surfactant solution consisting of 0.5~\%
volume fraction Dawn dishwashing liquid and 15 \% glycerol in
demineralized water (viscosity $\eta = 1.8 \pm 0.1$ mPa$\cdot$s
and surface tension $\sigma = 28 \pm 1$ mN/m) in a large circular reservoir (r = 190 mm) of depth 30 mm, see Fig.~\ref{setup}(a). A bidisperse (50:50 number ratio) bubble monolayer
is produced by flowing nitrogen through two syringe needles
immersed at fixed depth in the soapy solution. The resulting
bubbles of 1.8 $\pm$ 0.1 and 2.7 $\pm$ 0.1 mm diameter  are gently
mixed to produce a disordered bidisperse monolayer and are covered
with a 10 mm thick glass plate, see Fig.~\ref{setup}(a). The weighted average
bubble diameter $\langle d \rangle$ is 2.25 mm. We light the bubbles slanted from below 
and cover the bottom of the reservoir with a black plate, to enhance contrast.

The bubbles contact the top plate, see Fig.~\ref{setup}(b), which is completely wetted
by the soap solution, and the liquid fraction of the foam can be varied by varying the distance between glass plate and liquid
surface between 3 and 0.2 mm. This in itself is not a proper measure of $\phi$ since the relation between $\phi$ and the gap is strongly hysteretic --- $\phi$ depends not only on the gap, but also on an uncontrolled confining pressure. Therefore, we will determine $\phi$ from experimental images, in a procedure outlined below, see Fig.~\ref{setup}(c). We check that coalescence, segregation and coarsening are negligible.

\emph{Experimental protocol ---} A stepper motor is glued to the glass plate and is connected to an aluminum rod through a hole in the glass plate, see Fig.~\ref{setup}(b). We agitate the surfactant solution underneath the bubbles by driving the rod back and forth with an amplitude of 1 radian, its angular velocity alternating between +0.6  and -0.6 radians per second.
We emphasize that while agitating the bulk solution leads to strong mixing of the packing, we inject little enough energy to avoid bubble break-off, as evidenced by the absence of satellites.  After 4 oscillations we stop the motor, after which the packings slowly relax to a mechanically stable state. We probe the relaxation of packings at varying $\phi$ to determine the waiting time between agitation and image acquisition. To this end, we record sequences of images with a CCD camera and measure the variance of the intensity fluctuations of all pixels in difference images.

After waiting for this time, which is of the order of minutes, we record one image in the previously agitated region with a 6 megapixel photo-camera (Canon 20D). The image contains between 350 and 700 bubbles, depending on the packing fraction $\phi$. For various fixed gaps between liquid surface and glass plate we repeat this procedure 100 times and thus obtain 100 packings at roughly equal packing fraction. We visually inspect the resulting packings to be distinct in appearance, and it is these images that we analyse in the following.

\begin{figure}[ht]
\begin{center}
\includegraphics[width=\columnwidth]{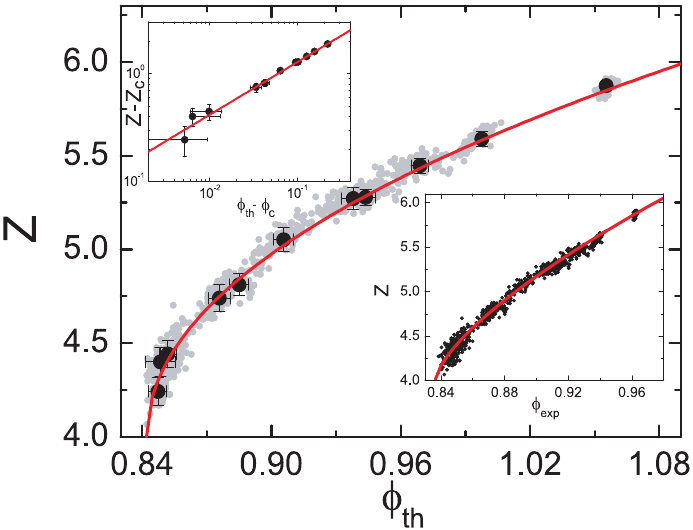}
\caption{(Color online)
Contact number Z of packings versus their packing fraction $\phi$. Grey scatter: data for every individual image. Circles: data averaged over experimental run at approximately constant packing fraction. Solid line is fit to ${\rm Z} = 4+ {\rm Z}_0(\phi-\phi_c)^{\beta}$, with  Z$_0 =4.02 \pm 0.20$, $\phi_c =0.842 \pm 0.002$ and $\beta=0.50 \pm 0.02$. Upper inset: same data on log-log scale. Lower inset: Z versus experimentally determined packing fraction $\phi_{exp}$. The fit has a power law exponent of $0.70$. } \label{images}
\end{center}
\end{figure}

\emph{Determining $\phi$ ---} We extract our crucial control parameter $\phi$ from the experimental images by advanced image analysis, see Fig.~\ref{images}(c): we first binarize the image, after which both the bubble centers and the interstices appear bright. We then remove the interstices by morpholog ical operations and end up with an image consisting of bright bubble centers. We dilate these centers and add up a negative of the original binary image --- in which the bubbles appear as bright rings --- to arrive at the final image, in which the bubbles are represented by bight discs against a black background. From this image we can readily calculate the area fraction $\phi$. Note that, in principle, the concept of packing fraction is problematic for a monolayer of three-dimensional bubbles. We choose our lighting of the bubbles such that the contacts between adjacent bubbles are optimally resolved. In other words, we image a slice from the packing where the bubbles are the broadest
and calculate a 2D packing fraction from this slice.

\emph{Scaling of Z with $\Delta \phi$} ---
We first determine the scaling of the  average contact number Z with $\phi$. To determine Z, we locate the center of mass of each bubble in the image, and after Delaunay triangulation and a subsequent removal of bond vectors for non-touching bubbles, we obtain the contact network of the bubbles in the image, from which we calculate Z,  see Fig.~\ref{setup}(c). 

Our results are presented Fig.~\ref{images}, where we plot the values for each distinct packing (grey dots), the average over all 100 images for each packing fraction (black circles) as well as a powerlaw fit of the form ${\rm Z} = 4+ {\rm Z}_0(\phi-\phi_c)^{\beta}$ (red solid line), where 4 is the contact number at isostaticity. The best fit gives us Z$_0 =4.02 \pm 0.20$, $\phi_c =0.842 \pm 0.002$ and $\beta=0.50 \pm 0.02$, in remarkable agreement with theoretical predictions by O'Hern et al and Durian. \cite{ohern, durian} who found Z$_0 =3.6 \pm 0.5$, $\phi_c =0.841 \pm 0.002$ and $\beta=0.49 \pm 0.03$.

Note however, that the range of packing fractions we can scan over, extends to a surprisingly large value of $\phi =1.06$. This is due to a striking discrepancy between the manner in which $\phi$ is calculated in simulations and in experiments. In simulations, the area or volume of spheres is fixed, and if one knows the number of particles in the periodic box, $\phi$ is readily calculated. In experiments, however, $\phi$ can only be inferred from experimental images. This difference results in the following: if particles overlap,  the overlapping area of the two particles is counted \emph{twice} in simulations, while it is only counted once in our experiment. This doubly counted area scales with the overlap $\xi$ as
$A_{ov} \sim \xi^{3/2}$, which stems from the fact that the deformed area scales as $r_c \times \xi = \sqrt{\xi}\times \xi$ \cite{lacasse}. Since $\xi \sim (\phi-\phi_c)$ \cite{vanhecke}, the conversion between a packing fraction extracted from a simulation $\phi_{th}$ and its experimentally accessible counterpart $\phi_{exp}$ should read:
\begin{equation}
\phi_{exp} = \phi_{th}-C(\phi_{th}-\phi_c)^{3/2}. \label{phiconv}
\end{equation}
We calculate both $\phi_{exp}$ and $\phi_{th}$
from numerically generated packings, and determine the pre-factor $C=0.95$. We then invert Eq.~(\ref{phiconv}) and calculate the $\phi_{th}$ corresponding to our $\phi_{exp}$.

When plotting our data against $\phi_{th}$ as in Fig.~\ref{images}, we excellently match simulations, while we find an apparent scaling exponent $\beta=0.70$ if we plot $Z$ as a function of the experimentally determined $\phi_{exp}$, see lower inset of Fig.~\ref{images}, owing to the nontrivial relation in $\Delta \phi$ between $\phi_{th}$ and $\phi_{exp}$.

We are not the the first to experimentally investigate the scaling of Z with $\phi$. Majmudar et al. \cite{majmudar} have extracted the same quantities from images of two-dimensional, frictional, photoelastic discs and compared these to predictions from simulations. From their data it appears the prefactor Z$_0 \approx 16$, inconsistent with simulations. Our results do allow for a direct comparison with frictionless jamming predictions, which can be seen from the excellent agreement between parameters.

\begin{figure}[ht]
\begin{center}
\includegraphics[width=\columnwidth]{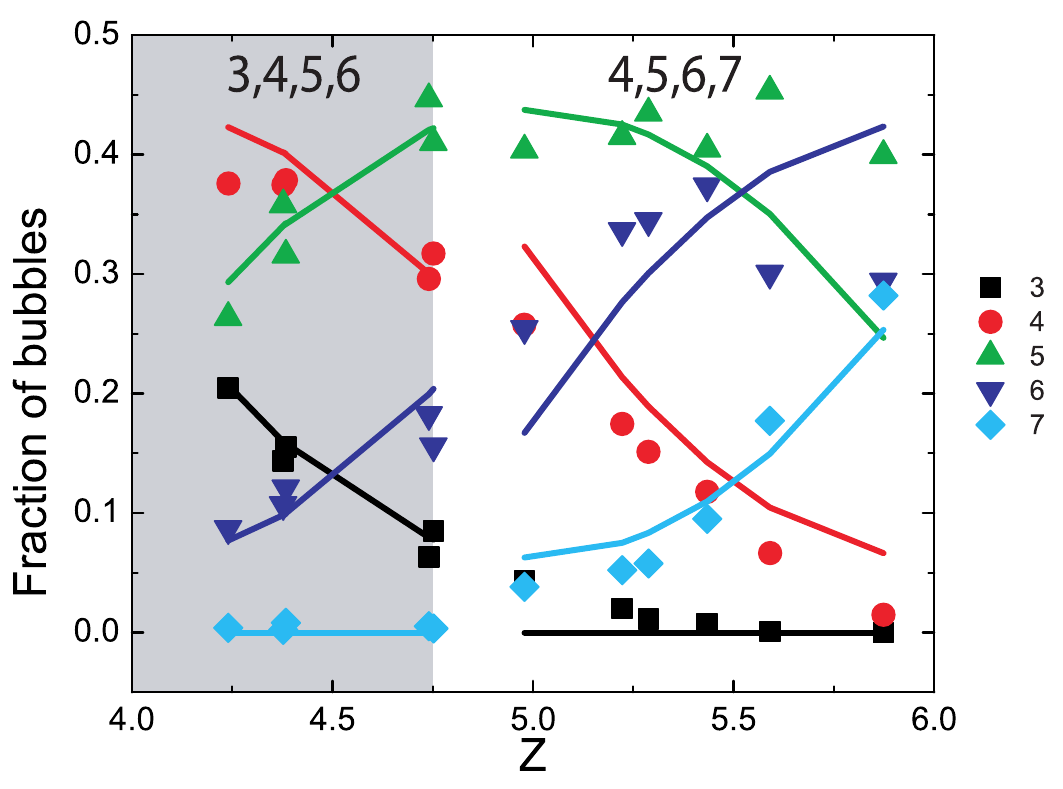}
\caption{(Color online)
Fractions of bubbles in the foam with $n$ contacts as a function of Z. Solid lines: solutions to Eqs.~(2-5) for the species listed at the top of the graph. 
} \label{fricfrac}
\end{center}
\end{figure}

\emph{Local contact fractions ---} Besides the average contact number per packing Z we can also extract the fraction $x_z$ of bubbles in each image that has $z$ contacts. We average these fractions over all images that correspond to a global packing fraction (and contact number Z) cf. the black circles in  Fig.~\ref{images}.
We plot these fractions versus the average Z in Fig.~\ref{fricfrac}: we see clear trends in the abundance of contacts at the particle level, to which we apply a very recent model \cite{henkes}.

This model predicts the fractions of 4 species $\{x_n,...,x_{n+3}\}$ in a packing, given the global Z and the variance $\sigma^2 = \sum^{n+3}_{i=n} x_i({\rm Z }-i)^2$. This constraint, together with the trivial normalization constraints $\sum^{n+3}_{i=n}x_i =1$, $\sum^{n+3}_{i=n}ix_i= {\rm Z }$ and the ill-understood, but empirically observed \footnote{both in \cite{henkes} and this work} constraint that the number of particles with odd and even contacts is equal, leads to a set of of equations, the solution of which is:
\begin{eqnarray}
x_n &=& \left(({\rm Z }-(n+2))^2+\sigma^2-1/2\right)/4 \label{eqna1} \\
x_{n+1} &=& \left(-({\rm Z }-(n+1))^2-\sigma^2+5/2\right)/4 \label{eqna2} \\
x_{n+2} &=& \left(-({\rm Z }-(n+2))^2-\sigma^2+5/2\right)/4 \label{eqna3} \\
x_{n+3} &=& \left(({\rm Z }-(n+1))^2+\sigma^2-1/2\right)/4 \label{eqna4}
\end{eqnarray} 

Since we know Z and $\sigma^2 (=0.75)$ from the data we can obtain the fractions $x_i$ without any free parameters. However, we measure non-negligible fractions of not 4, but 5 species. We therefore apply the model for $n=3$ to $4<~$Z$~<4.75$, where $x_7 \approx 0$ and for $n=4$ to $4.97<~$Z$~<6$ where $x_3 \approx 0$. We obtain good agreement between data and theory, see Fig.~\ref{fricfrac}, for both ranges of  validity.  At high Z, the bidisperse nature of our packing is visible: we observe an excess of both particles with 5 contacts and with 7, while particles with 6 contacs are underrepresented. We have observed that large bubbles carry the majority of 7's, while small bubbles mostly have 5 contacts at these values of Z (data not shown). This is natural in bidisperse packings, as this occurs whenever a large and a small bubble contact. It thus indicates an absence of crystalline order, which would lead to an increase of 6's.
\emph{Area distributions ---}
We now turn to tesselations of our foam packings. These tesselations yield two important "connectors" between local geometry and global response. Firstly one can readily extract the distribution of available area per bubble $p(A)$ \cite{lechenaultII,schroeder,kumaran,slotterback,kudrolli, cheng}, which serves as the multiplicity in a thermodynamical description of granular materials. Secondly, the size of the contact area's between bubbles can be measured, from which the distribution of forces $p(f)$ can be extracted which is needed as an input into any prediction of the mechanical properties of packings. 

The thermodynamical description of granular materials, as introduced by Edwards and Oakeshott \cite{edwards} translates the concepts underpinning equilibrium thermodynamics to conglomerates of a-thermal particles such as bubbles. The volume $V$ (in 2D the area $A$) takes on the role of energy, while a compactivity $\chi$ replaces temperature.
\begin{figure*}[ht]
\begin{center}
\includegraphics[width=\linewidth]{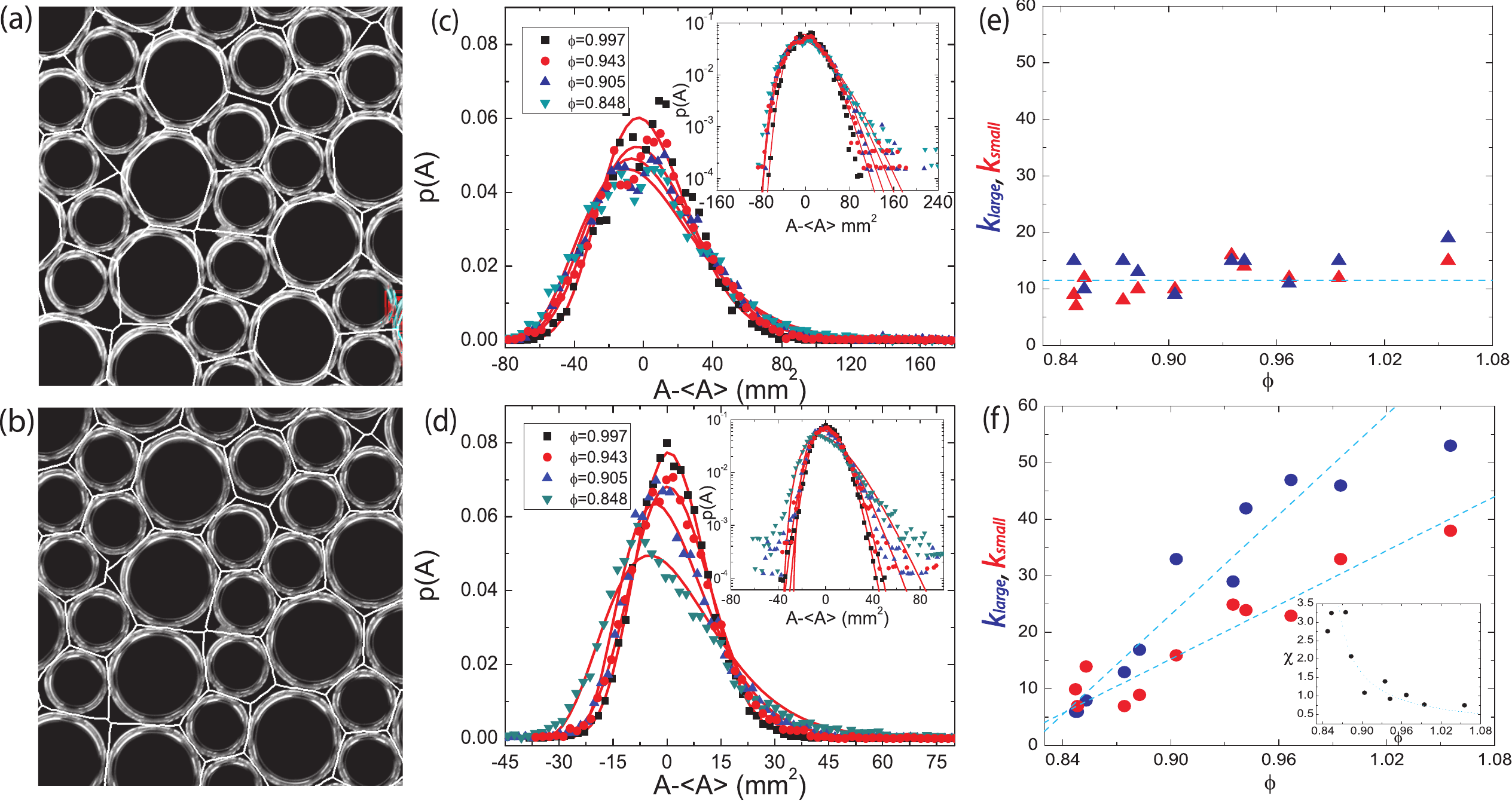}
\caption{(Color online)
(a) Experimental image  with Voronoi tesselation of the centers of mass of the bubbles. Note the intersection of bubbles by the Voronoi cell edges. (b) Navigation map tesselation of same image: the cell edges do not intersect the bubbles. (c)  Area distributions from Voronoi tesselation for 4 different packing fractions, as stated in legend. Solid lines are fits to Gamma distribution Eq.~\ref{astevor}. Little trend with $\phi$ is seen. (d) Area distributions from Navigation map tesselation for the same packing fractions. Solid lines are fits to Gamma distribution Eq.~\ref{astevor}. Near $\phi_J$, broad tails develop.
(e) $k$ extracted from fits to Voronoi cell distributions. No trend with $\phi$ can be seen. (f) $k$ extracted from fits to Navigation map distributions. A strong variation of $k$ with $\phi$ is visible. The inset shows the compactivity $\chi = (\langle A \rangle - A_{min})/k$, which increases towards point J as $k$ decreases. } \label{navvor}
\end{center}
\end{figure*}

Aste and Di Matteo have derived the form of the distribution function $p(A,k)$ with such an approach \cite{aste}:
\begin{equation}
p(A,k) = \frac{k^k}{(k-1)!}\frac{(A -A_{min})^{k-1}}{(\langle A \rangle -A_{min})^{k}} \exp\left(k \frac{A -A_{min}}{\langle A \rangle -A_{min}}\right), \label{astevor}
\end{equation}
\par
with $k$ a shape parameter --- which has been found to take on one universal value in granular packings, namely $k=12$ \cite{schroeder}, $A_{min}$ the minimum available area per bubble and $\langle A \rangle$ the mean.

For deformable particles, it is not immediately obvious how the areas $A$ should be computed. Here we calculate $p(A)$ from our experimental images for each packing fraction both from a Voronoi and a navigation map tesselation, which are detailed below.  Upon fitting Eq.~(\ref{astevor}) to our data we obtain excellent matches to all distributions measured in both ways, but we find strongly nonuniversal behaviour in the parameter $k$  for the navigation map method which we argue to be the correct one.

We measure the probability distribution of free areas $p(\emph{A})$
by calculating the Voronoi area distribution of the grid of points that represent the centers of mass of the bubbles. However, Voronoi cell edges do in general not respect the bubble perimeter, see Fig.~\ref{navvor}(a) and thus the Voronoi cell does not represent the free area per bubble. For hard spherical objects one can get around this problem by weighting the grid points according to the sphere radius (Voronoi-Laguerre tessellation) \cite{lechenaultII}, however, in our experiment, the bubbles are not only bidisperse, but in general also deformed and the flattened contacts can be curved.

To fully take the effects of both deformations and bidispersity into account, we calculate what is called the navigation map \cite{medvedev, corwin}. In this method, we assign the interstitial area to those bubbles whose perimeter is closest. The result is shown in Fig.~\ref{navvor}(b): we obtain free areas per bubble that respect the bubble edges and follow the curvature of the contacts.

With both methods we obtain bimodel distributions for $A$, which we split according to the size of the bubbles. Distributions
for the larger bubbles are shown in Fig.~\ref{navvor}(c,d) for Voronoi and navigation map tesselations respectively. Distributions for the smaller bubbles behave the same. For the Voronoi tesselation we find that the shape of the distributions is roughly independent of the packing fraction, with all distributions being slightly skewed (see inset of Fig.~\ref{navvor}(c)), while for the navigation map tesselation we find that an excess of large available area develops near jamming, leading to strongly asymmetric distributions.

We quantify these observations by fitting Eq.~(\ref{astevor}) to our experimental distributions. We treat $k$ as a fit parameter and extract $A_{min}$ from the data: it is the minimal hexagon that can enclose a bubble at a given packing fraction. Results are plotted in Fig.~\ref{navvor}(e,f): the nearly constant shape of the Voronoi distribution is reflected by the near constant value of $k$ we extract, with $<k>=13.1$ for large bubbles and $<k>=11.4$ for small bubbles when averaged over all values of $\phi$. On the other hand, we observe a systematic trend in $k$ for the navigation map distributions, the nearly Gaussian distributions at high $\phi$ can be fit when $k \sim 50$, but $k$ systematically decreases with decreasing packing fraction $\phi$ to a value of 6 near jamming.

We thus observe that $p(A)$ strongly depends on the particular tesselation we use to extract it. For Voronoi tesselations we find a $k$ which is remarkably close to the  value found by Aste and coworkers in hard-sphere packings, while for the navigation map we find a strong variation of $k$ with $\phi$, that reflects the increasing amount of excess available area per bubble with decreasing packing fraction, which is also reflected by an increasing compactivity $\chi = (\langle A \rangle - A_{min})/k$ \cite{aste}, see inset of Fig.~\ref{navvor}(f). Since the navigation map tesselation respects bubble edges, we believe this tesselation to be more physically appropriate, and we see the decrease of $k$ and the associated broadening of the tail of $p(A)$ as a signature of the approach of point J, or, equivalently, the recovery of hard-sphere behaviour.

\begin{figure*}[ht]
\begin{center}
\includegraphics[width=\linewidth]{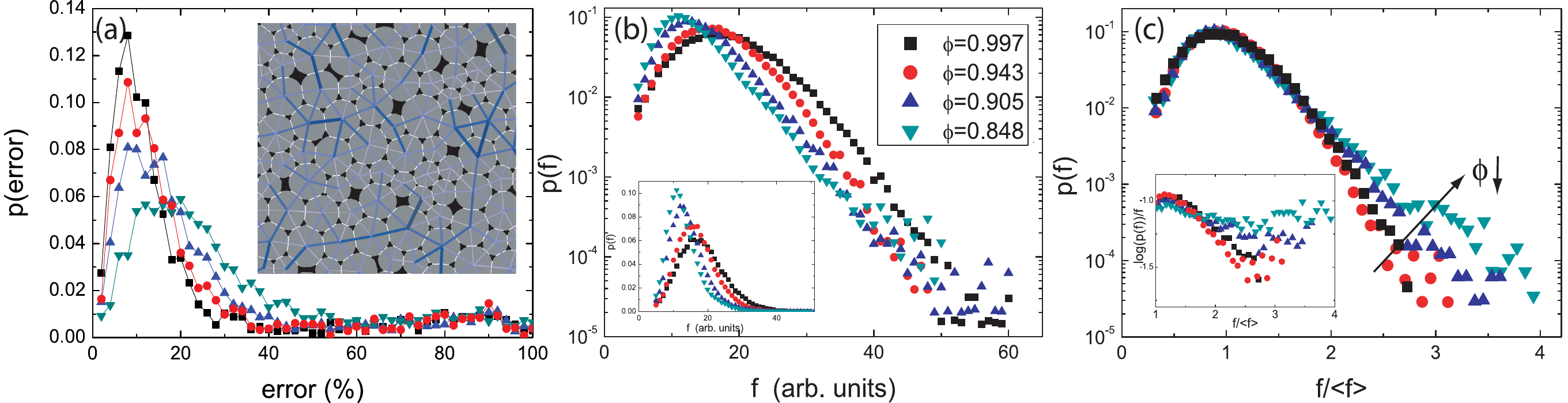}
\caption{(Color online)
For the same $\phi$ as in Fig.~\ref{navvor} we calculate (a) Probability distribution of relative error in force balance on each bubble: for packings closer to point J, the error increases. Inset: experimental image with force network overplotted: flattened contacts in white, and magnitude of the forces indicated by thickness of vector and darkness. (b) Force distributions, rescaled by the mean $f$ of all configurations: the peak shifts to smaller forces as one approaches point J (see inset). (c) Force distributions, for each configuration we rescale its $p(f)$ by the mean. The average now exhibits steeper than exponential decay.} \label{p(f)}
\end{center}
\end{figure*}

\emph{The force distribution p(f) ---} By construction, the navigation map bisects touching bubbles at their contact area. As a result, we can
extract the radius $r_c$ of the deformed contacts between bubbles. Since the contact force $f_{12}$ between two bubbles is linear in the deformation, it is related to $r_c$ as \cite{emul}
\begin{equation}
f_{12} = \pi r_c^2\left( \frac{R_1+R_2}{R_1 R_2} \right) 
\label{princen}
\end{equation}
with $R_i$ the radius of bubble $i$. We can thus extract the force distribution $p(f)$ of interbubble contact forces. An experimental image with the contacting facets in white and the extracted contact forces in blue is shown in Fig.~\ref{p(f)}(a). 

The forces are oriented along the bond vector between bubbles and thus we can check if the forces balance on each bubble. To check this, we plot the relative error $\| \sum {\bf f} \| /\sum \| {\bf f} \| $ in Fig.~\ref{p(f)}(a). We see that the error is about 10 \% for the densest packings, and increases for packings that are closer to jamming.
 
In Fig.~\ref{p(f)}(b) we plot, for various $\phi$, the $p(f)$ of all forces from all images at that packing fraction. We convert $r_c$ to forces using the blob radii $R_i$ we obtain after binarizing and removing the interstices. These radii are proportional to, but smaller than the actual bubble radii. We therefore plot the force distributions in arbitrary units. The distributions are peaked and exhibit broad tails, reflecting the heterogeneity in the forces. Their shapes are similar to those found in grains and emulsions \cite{emul,grain},  In the inset of Fig.~\ref{p(f)}(b) that the peak of the distribution shifts towards lower forces as we approach point J, in accordance with \cite{ohern2}.

To obtain more information on the shape of the tail of $p(f)$, we now rescale each $p(f)$ from an individual image by its mean and average these distributions for each $\phi$. Just adding up the force distributions for each frame, as was done for Fig.~\ref{p(f)}(b), washes out any signature of Gaussian tails \cite{ohern2}. The result is shown in Fig.~\ref{p(f)}(c): we now observe changes in the shape of the tail away from jamming --- with increasing compression the decay of the tail seems to become steeper than exponential. This is illustrated by the inset of Fig.~\ref{p(f)}(c), where we plot $ln(p(f))/f$ which tends to a constant for exponential tails and decreases for faster than exponential decay. We thus provide further evidence that one can find steeper than exponential decay in $p(f)$, although we do not have sufficient statistics to conclude that the tails cross over to Gaussian tails away from jamming.

{\bf Conclusion.} We have experimentally investigated the behavior of soft frictionless discs near the jamming point at zero stress and temperature, by sampling distinct packings of foam bubbles. Experimental data of systems near jamming are still scarce and we hope that the various measures that we have extracted can help to decide between various competing theoretical descriptions of jammed matter.

\begin{acknowledgments}
The authors wish to thank Jeroen Mesman and Joshua Dijksman for technical assistance. The authors thank Zorana Zeravcic for providing numerically generated packings and Wim van Saarloos and Brian Tighe for useful suggestions.
GK acknowledges support from physics foundation FOM, and MvH
acknowledges support from NWO/VIDI.
\end{acknowledgments}

\end{document}